\documentclass[aps,prl,twocolumn,showpacs,floatfix]{revtex4-2}
\usepackage[utf8]{inputenc}
\usepackage[T1]{fontenc}
\usepackage{float} 
\usepackage{color}  
\usepackage{graphicx}
\usepackage{amsmath}
\usepackage{amssymb}
\usepackage{bbm}
\usepackage{indentfirst}
\usepackage{dcolumn}
\usepackage{soul}
\usepackage{mathrsfs}
\usepackage{xcolor}
\usepackage{ulem}

\usepackage{relsize}
\usepackage{amsmath}
\usepackage{graphicx}
\usepackage{float}

\usepackage{color}
\makeatletter

\usepackage{xcolor}
\usepackage{soul}
\usepackage{soulpos}

\usepackage{dcolumn}
\usepackage{bm}
\usepackage{subfigure}
\usepackage{helvet}
 \usepackage{color}
\graphicspath{{figures/}}

\usepackage{tcolorbox}

\usepackage[colorlinks=true,linkcolor=blue,citecolor=blue,urlcolor=blue]{hyperref}

\begin{document}

\title{Exchange-enhancement of the ultrafast magnetic order dynamics in antiferromagnets}

\author{Florian Jakobs}
\affiliation{Dahlem Center for Complex Quantum Systems and Fachbereich Physik, Freie Universit\"{a}t Berlin, 14195 Berlin, Germany}
\author{Unai Atxitia}
\email{unai.atxitia@fu-berlin.de.}
\affiliation{Dahlem Center for Complex Quantum Systems and Fachbereich Physik, Freie Universit\"{a}t Berlin, 14195 Berlin, Germany}
\date{\today}

\begin{abstract}
We theoretically demonstrate that the ultrafast magnetic order dynamics in antiferromagnets is exchange-enhanced in comparison to their ferromagnetic counterparts. We provide an equation of motion for the magnetic order dynamics validated by computer simulations using atomistic spin dynamics methods. The exchange of angular momentum between sublattices speeds up the dynamics in antiferromagnets, a process absent in ferromagnets.  
\end{abstract}

\maketitle

Ultrafast optical control of the magnetization promises faster data processing and storage \cite{Beaurepaire1996,KirilyukRMP2010,Koopmans2010,RaduNature2011,OstlerNatComm2012,BattiatoPRL2013}. 
Antiferromagnets (AFMs) show advantages over ferromagnets (FMs), such as faster magnetization dynamics  \cite{Kimel2004,Wadley2016,Baltz2018,JungwirthNatNanotech2016,Olejnik2018,RozsaPRB2019,Kaspar2020}. 
In AFMs, the frequency of the magnetic oscillations around the anisotropy field in FMs ($\omega_{\rm{fm}} \sim H_{\rm{A}}$) are exchange-enhanced by the antiferromagnetic coupling ($H_{\rm{E}}$) between the spins at different sublattices, leading to a higher oscillation frequency in AFMs, ($\omega_{\rm{afm}} \sim \sqrt{H_{\rm{E}} H_{\rm{A}}}$), orders of magnitude higher than in FMs\cite{Kittel1951,Kampfrath2011}.
Femtosecond laser photo-excitation can induce subpicosecond magnetic order quenching in both FMs and AFMs \cite{Beaurepaire1996,Fiebig2008,Thielemann-Kuhn2017}.  
The speed of ultrafast quenching of the magnetic order is determined by the strength of the exchange interaction ($1/\tau_{\rm{fm}} \sim \alpha_{\rm{fm}} H_{\rm{E}}$), and the FM damping, $\alpha_{\rm{fm}}$ \cite{Atxitia2011}. 
This raises the fundamental question of whether the ultrafast magnetic order dynamics in AFMs is exchange-enhanced with respect to its FM counterpart.
We find that the AFM magnetic order responds 
faster than the FM one to a sudden temperature change due to the exchange-enhancement of the effective AFM damping, $1/\tau_{\rm{afm}} \sim \alpha_{\rm{afm}} H_{\rm{E}}$. We show that, contrary to FMs, the effective AFM damping depends on the number of neighbours to which spins are exchange coupled. Thus, low dimension magnets, such as 1D and 2D magnets\cite{Genome22}, show a more pronounced speed up, while for lattices with higher coordination number the exchange-enhancement reduces.  
As the system approaches the critical temperature, both the FM and AFM present a critical slow down of the relaxation process, however, the AFM critical exponent is smaller than the FM one. 
In the temperature dominated regime $T \gg T_{\rm{c}}$, our model predicts intrinsically different relaxation dynamics for AFMs and FMs. This scenario corresponds to experiments using powerful femtosecond laser pulses.
For FMs, magnetic order quenching slows down as the magnetization reduces, while for AFMs speeds up.
We demonstrate the validity of our model by direct comparison to computer simulations using atomistic spin dynamics within an atomistic spin dynamics (ASD) model.

Evidence of exchange-enhancement of the ultrafast magnetization dynamics in AFMs is scarce due to the difficulties to conduct a systematic comparison on the same system presenting FM and AFM magnetic order. 
Studies in rare-earth Dy using femtosecond time-resolved resonant magnetic x-ray diffraction have measured the dynamics of its FM and AFM-spin-helix states. 
These investigations have shown that the dynamics of the order parameter in the AFM phase is faster than in the FM phase  \cite{Thielemann-Kuhn2017}.
Laser induced ultrafast magnetization dynamics in FMs have been modeled using computer simulations based on different approaches, from ASD models to macroscopic phenomenological approaches \cite{Beaurepaire1996,Koopmans2010,Atxitia2011}.  
Within these approaches magnetization dynamics are explained on thermodynamic grounds. When the temperature of the heat-bath is rapidly modified, the magnetic order changes according to  $\dot{M} \sim \alpha_{\rm{fm}} H$,  driven by an effective field $H=-\partial F/\partial M$ at a rate $\alpha_{\rm{fm}}$, towards minimal free energy values, $F(M)$ \cite{Baryakhtar2013}.  
The thermodynamic argument explains demagnetization and magnetization recovery when the system temperature increases and decreases, respectively. 
It can be expected to hold for other magnetic structures, such as antiferromagnets, as well. In AFMs however, an additional channel for angular momentum dissipation opens, by direct exchange of angular momentum between sublattices. For a two sublattice AFM, the dynamics of sublattice $a$ can be expressed as : $\dot{M}_a \sim  \alpha_a H_a +\alpha_{\rm{ex}}(H_a-H_b)$, where $\alpha_{\rm{ex}}$ represents the rate of interatomic transfer of angular momentum between sublattices $a$ and $b$.
In the simplest AFM case, both sublattices are equivalent, such that $H_a=-H_b$ is a valid approximation, leading to $\dot{M}_a \sim (\alpha_a + 2 \alpha_{\rm{ex}}) H_a = \alpha_{\rm{afm}} H_a $, where $\alpha_a$ and $\alpha_{\rm{ex}}$ are the Onsager coefficients describing exchange and relativistic relaxations. For FM and AFM systems defined by the same parameters, $\alpha_{\rm{afm}}>\alpha_{\rm{fm}}$, and consequently the AFM is faster than the FM. 
In the present work, starting from an ASD model, we derive an -- so far unknown -- expression for the exchange-enhancement of the relaxation parameter, $\alpha_{\rm{afm}}$ in AFMs.

The dynamics of the magnetic order parameter (in FMs and AFMs) are calculated within the framework of a classical, atomistic spin model. The Hamiltonian reads 
\begin{eqnarray}
\mathcal{H}=-\frac{J}{2}\sum_{\langle i,j \rangle }\mathbf{s}_{i}\mathbf{s}_{j}
-d_z\sum_{i} \left(s_{i}^{z}\right)^{2}.
\label{eq:Hamiltonian}
\end{eqnarray}
The unit vectors, $\mathbf{s}_i= \boldsymbol{\mu}_i/\mu_{\rm{at}}$, represent the normalized magnetic moment of the lattice site $i$ with magnetic moment $\mu_{\rm{at}}=\mu_{\rm{B}}$. The first term describes nearest neighbors exchange coupling, with $J= \pm 3.450 \times 10^{-21}$ J for the FM (+) and AFM(-).
The second term represents the uniaxial anisotropy, with $d_{z} = 1 \times 10^{-22}$ J for both AFM and FM. 
The dynamics at finite temperatures are described by the stochastic Landau-Lifshitz-Gilbert (s-LLG) equation,
 \begin{eqnarray}  
           \frac{d\mathbf{s}_i}{dt} =
              - \frac{|\gamma|}{(1+\lambda^2)}  \mathbf{s}_i  \times  \left[ \mathbf{H}_i 
             -  \lambda \;   \left( \mathbf{s}_i  \times \mathbf{H}_i  \right) \right].
             \label{eq:LLG}
\end{eqnarray}
Here, $\gamma$ is the gyromagnetic ratio. 
The first term represents a precession of the magnetic moments around an effective field $\mathbf{H}_i = - (1/\mu_{\rm{at}} )(\partial \mathcal{H}/\partial \mathbf{s}_i)$, while the second term represents the transverse relaxation.  A phenomenological  damping constant $\lambda$ defines the rate of the relaxation. 
In order to include the effects of finite temperature,
 we couple the spin system to a Langevin thermostat which adds an effective field-like stochastic term $ \boldsymbol{\zeta}_i$ to the effective field  with white noise properties~\cite{Atxitia2009}.

By using ASD simulations, we first demonstrate the existence of exchange-enhancement on the AFM dynamics in realistic conditions, similar to experiments. A high temperature regime, $T \gg T_{\rm{c}}$ can be accessed by suddenly heating the electron system using a femtosecond laser pulse (Fig. \ref{fig:Fig1} (a)) (see more details in supplementary material Sec. S1). On the timescale of 100 fs, the electron temperature will rise far above $T_{\rm{c}}$.

The magnetic system responds to this temperature change by reducing its magnetic order on similar time scales.  The electron-phonon coupling allows energy transfer from the hot electrons to the lattice in the time scale of only a couple of picoseconds. This allows for the investigation not only of the magnetic order quenching but also its recovery.  
Figure \ref{fig:Fig1} (b) shows that the demagnetization in the AFM is larger than in FM, owning to a faster response when excited by the same temperature profile (Fig. \ref{fig:Fig1} (a)).
The magnetic order recovery of the AFM is on the same time scale as the electron-phonon temperature relaxation time, while the FM relaxes over longer time scales.

\begin{figure}[h]
\centering
\includegraphics[width=\columnwidth]{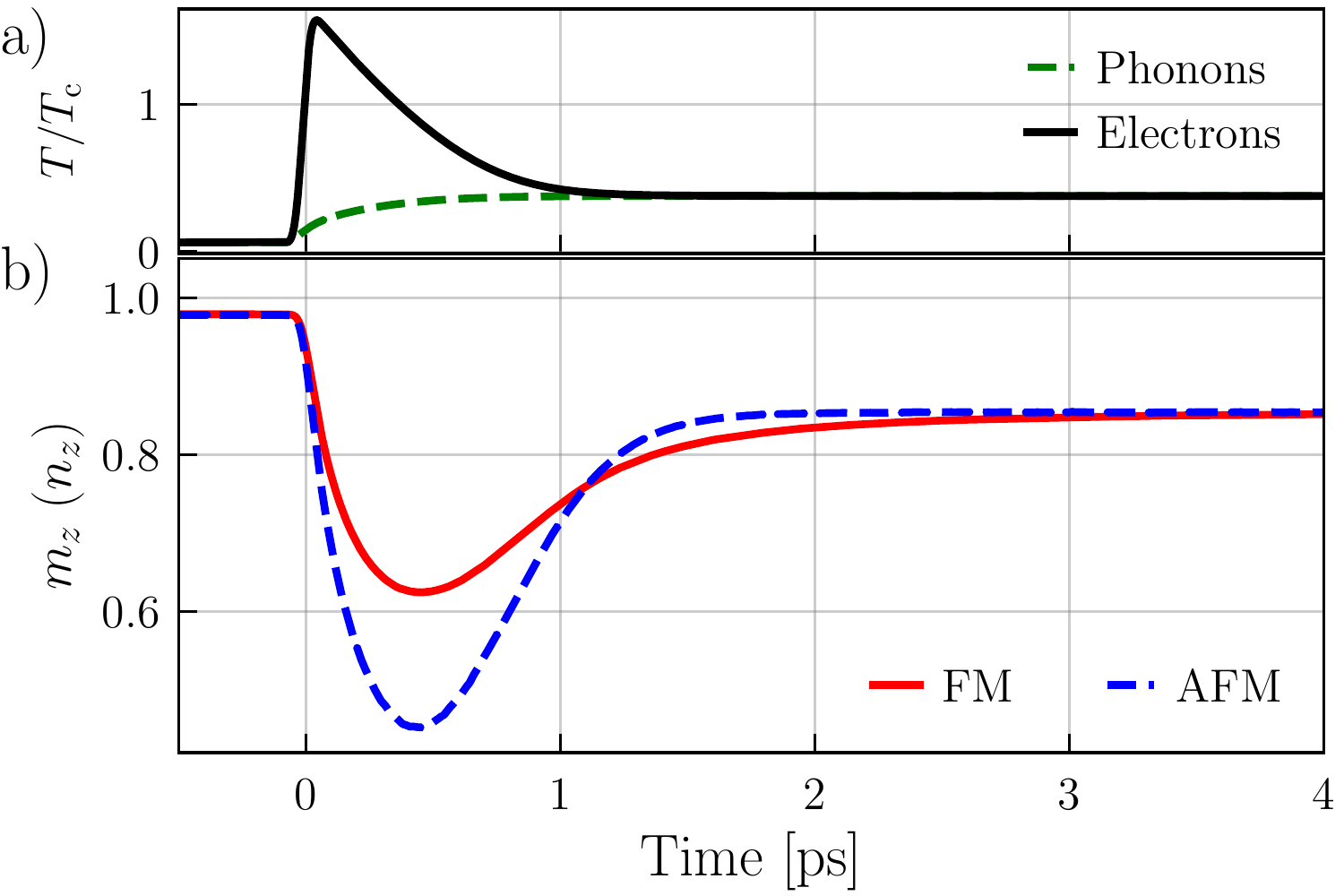}\caption{(a) Electron and phonon temperature dynamics after an excitation by a 50 fs laser at $t=0$. (b) The magnetic order dynamics of a FM, $m_z$ (red solid line), and an AFM,  $n_z$ (blue dashed line) as a response to the electron temperature dynamics in (a).}
\label{fig:Fig1}
\end{figure}
The exchange-enhancement of the AFM magnetic damping is at the  origin of this speed up as we shall demonstrate.

Building upon the described atomistic spin model, the non-equilibrium macroscopic magnetization dynamics of the sublattice $m_a=\langle s_a \rangle$  can be described by \cite{MentinkPRL2012}:
\begin{equation}
\frac{1}{\gamma}\frac{d m_a}{d t} =  \alpha_{a}  H_a + \alpha_{\rm{ex}} ( H_a -  H_b )
\label{eq:long-LLB-1-ex}
\end{equation}
where $a \neq b$.
The macroscopic damping parameter in Eq. \eqref{eq:long-LLB-1-ex} is defined as, $\alpha_a = 2\lambda L(\xi_a)/\xi_a$.
Here, $L(\xi_a)$ stands for the Langevin function, with the argument $\xi_a = \beta \mu_a H_a^{\rm{MFA}}$ and $\beta=1/k_{\rm{B}} T$ \cite{Garanin1997}.
In the exchange approximation, the MFA field acting on sublattice $a$ is $\mu_a H_a^{\rm{MFA}} = J_0 m_b$,  here $J_0=zJ$, where $z$ is the number of nearest neighbours of spins of type $b$.
 Moreover, in the exchange approximation, one can fairly assume that $m_a=m_b=m$, and therefore $\alpha_a=\alpha_b$.  
Under these assumptions, the exchange relaxation parameter can be written as $\alpha_{\rm{ex}}= 4\alpha_a/(zm)$ \cite{Jakobs2022arxiv}. 
One can recover the equation of motion for the FM case for $\alpha_{\rm{ex}}=0$ (Eq. \eqref{eq:long-LLB-1-ex}), in that case, $\alpha_{\rm{fm}}=\alpha_a$.
The non-equilibrium effective fields are given by 
\begin{equation}
H_a = \frac{(m_a-m_{0,a})}{\mu
_a\beta L'(\xi_a)}.
\label{eq:muaHa}
\end{equation}
where, $L'(\xi) = dL/d\xi$ and $m_{0,a}=L(\xi_a)$ \cite{Garanin1997,Atxitia2012}. 

 For the two sublattice AFM considered here, $H_a = - H_b = H_n$, where $H_n= (n-n_0)/\mu_{a}\beta L'(\xi)$. 
 It follows that the dynamics of the N\'{e}el order parameter is given by
\begin{equation}
\frac{1}{\gamma}\frac{d n}{d t} =  \alpha_{\rm{afm}}   H_n.
\label{eq:neel-dynamics}
\end{equation}
This demonstrates that the origin of exchange-enhancement of the AFM dynamics can be traced back to the effective AFM damping parameter,
\begin{equation}
\alpha_{\rm{afm}} = \alpha_{\rm{fm}} \left(1+\frac{4}{z|n|}\right).
\label{eq:effective-damping-AFM}
\end{equation}
\begin{figure}[t]
\centering
\includegraphics[width=\columnwidth]{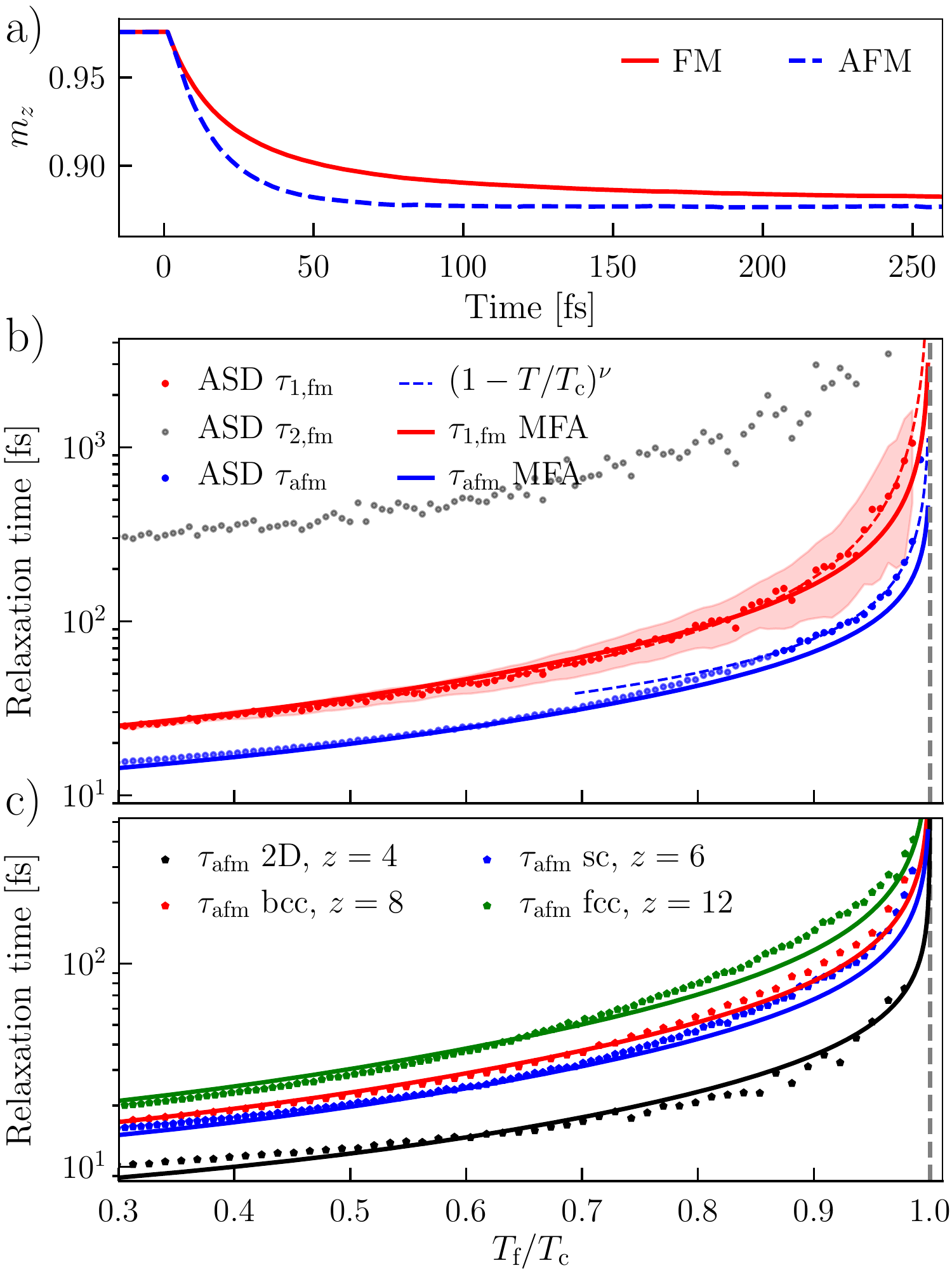}
\caption{(a) Magnetic order dynamics of a FM and an AFM after a step-like temperature increase, $T_{\rm{i}}/T_{\rm{c}}=0.07$ and 
final temperature $T_{\rm{f}}/T_{\rm{c}}=0.33$.
(b) Magnetic order relaxation time in FMs and AFMs as a function of the reduced temperature $T_{\rm{f}}/T_{\rm{c}}$.
Symbols correspond to ASD simulations, solid lines to MFA, and dashed lines are a fit of the scaling law, $(1-T/T_{\rm{N}})^{\nu}$ with $\nu=-1.017(7)$ for $\tau_{\rm{1,fm}}$ and $\nu=-0.64(2)$ for $\tau_{\rm{afm}}$. The light red area around the FM data indicates the statistical uncertainty  coming from over 50 simulations.
For AFMs is not shown since it is around 2-4\% at maximum close to $T_{\rm{N}}$.
(c) The relaxation time for different lattice structures (sc, bcc, fcc and a 2D square), each with different number $z$ of exchange coupled neighbouring spins. Symbols correspond to ASD simulations and lines to MFA.}
\label{fig:relaxtimeEq}
\end{figure}    
We first address the differences and similarities between AFM and FM near thermal equilibrium, where the non-equilibrium fields can be cast into Landau-like expressions \cite{MentinkPRL2012,Atxitia2012}, $H_n = (\mu_{\rm{at}}/2\widetilde{\chi}_{\|})\delta n^2/n_{e}^2$. Here $\widetilde{\chi}_{\|}$ is the longitudinal susceptibility of the N\'{e}el vector at zero field
 \begin{eqnarray}  
           \widetilde{\chi}_{\|} = 
                  \frac{\mu_{\rm{at}}}{J_0} \frac{\beta J_0 L'}{1-\beta |J_0| L'}.
             \label{eq:FMlongitudinalSusceptibility}
\end{eqnarray}
For small deviations $\delta n_e$ of the order parameter $n$, from equilibrium ($\delta n_e \ll n_e$), Eq. \eqref{eq:neel-dynamics} can be expanded around the equilibrium state $n_e$. The resulting linear equation in $n$ can be easily solved analytically as an exponential decay,  with relaxation time given by $\tau_{\rm{afm}} = \widetilde{\chi}_{\|}/\gamma \alpha^{e}_{\rm{afm}}$, where $\alpha^{e}_{\rm{afm}}$ is calculated for $n=n_e$. 
At low temperature, where $n_e \approx 1$ the ratio between AFM and FM relaxation time is just given by $\alpha_{\rm{afm}}/\alpha_{\rm{fm}}= 1+4/zn_e$, which equals to $1+4/z$ (5/3 for simple cubic) at strictly $T=0$ K. In striking contrast to the FMs, the effective AFM damping depends on the number of nearest neighbours $z$. For the MFA limit, $z\rightarrow{\infty}$, the exchange-enhancement vanishes ($\alpha_{\rm{afm}}=\alpha_{\rm{fm}}$) whereas for systems with low coordination number increases, for example a spin chain with $z=2$, $\alpha_{\rm{afm}}=3\alpha_{\rm{fm}}$. 
Another relevant example would be the metallic antiferromagnet Mn$_2$Au \cite{barthem2013revealing,roy2016robust,meinert2018electrical}, the most promising candidate for future spintronic and memory applications, in which each Mn spin is antiferromagnetically exchange coupled to five neighbours, and thus a speed up of a factor  $1+4/5=1.8$ is expected. 
At temperatures approaching $T_{\rm{c(N)}}$, the order parameter reduces, $n_e \approx 0$, and consequently, $\alpha_{\rm{afm}}/\alpha_{\rm{fm}} \sim 1/n_e$ diverges.  In the MFA, close to $T_{\rm{c(N)}}$, the  order parameter at equilibrium scales with temperature as $n_e \sim (1-T/T_{\rm{c(N)}})^{1/2}$. Hence the critical behaviour of the AFM damping parameter is $\alpha_{\rm{afm}} \sim \left(1-T/T_{\rm{N}}\right)^{-1/2}$.
While the longitudinal susceptibility increases with temperature up to the critical temperature where it diverges, $\widetilde{\chi}_{\|} \approx (1-T/T_{\rm{c}(\rm{N})})^{-1}$.
Thus, the relaxation time in AFMs scales as $\tau_{\rm{afm}} \sim (1-T/T_{\rm{N}})^{-1/2}$, which differs from the scaling for FMs $\tau_{\rm{fm}} \sim \left(1-T/T_{\rm{c}}\right)^{-1}$. Although both AFMs and FMs show the so-called critical slow down near $T_{\rm{c(N)}}$, the effect of the  exchange-enhancement of the AFM dynamics is to lower the critical exponent. Since, in general, the MFA scaling laws are known to differ from the actual critical scaling exponents, $\tau \sim \left(1-T/T_{\rm{c}}\right)^{-\nu}$, in the following we conduct ASD computer simulations to verify qualitatively these theoretical predictions,  i) to find quantitatively the  critical exponents, $\nu$, for AFMs and FMs, and ii) to demonstrate the dependence on the number of exchange links between spins of the relaxation time in AFMs. 
To do so, we compute the relaxation time under the same conditions, namely, for small deviations from equilibrium, $ \delta n_e /n_e \ll 1$.
This is achieved by applying a step-like temperature increase, $\Delta T=T_{\rm{f}}-T_{\rm{i}}$, such that $\delta n = n_e(T_{\rm{i}})-n_e(T_{\rm{f}})=0.1 n_e(T_{\rm{i}})$ for all initial/final temperatures $T_{\rm{i}}$/$T_{\rm{f}}$.
We clarify for direct comparison of the ASD simulations to analytical estimations, the parameters have to be calculated for the final temperature, $n_e(T_{\rm{f}})$.
An example of such magnetization dynamics for the $z-$component of the order parameter for FM and AFM orderings for a simple cubic lattice ($z=6$) are shown in Fig.~\ref{fig:relaxtimeEq}~(a). 
We find that for FMs the relaxation dynamics is defined by two characteristic times, $\tau_{1,\rm{fm}}$ and $\tau_{2,\rm{fm}}$, associated to a fast and a slow relaxation process, respectively and for the AFM a single $\tau_{\rm{afm}}$ is enough to describe the demagnetization process.
Fig.~\ref{fig:relaxtimeEq}~(b) shows the relaxation times $\tau_{1,\rm{fm}}$, $\tau_{2,\rm{fm}}$ and $\tau_{\rm{afm}}$ as function of the reduced temperature $T_{\rm{f}}/T_{\rm{c}}$ in comparison to the MFA analytical expression derived from Eq.~\eqref{eq:neel-dynamics}.
We note that the values for the relaxation time in FMs in Fig.~\ref{fig:relaxtimeEq}~(b) are obtained from averaging over 50 individual ASD simulations.
Interestingly, we find that for all temperatures the slow relaxation time  $\tau_{2,\rm{fm}}$  is related to fast one $\tau_{1,\rm{fm}}$ as $\tau_{2,\rm{fm}} = 12 \tau_{1,\rm{fm}}$. 
The faster time decay is related to the relaxation of the magnetic order, while the slower one with the relaxation of short-wavelength spin waves\cite{Baryakhtar2013}. 
Differently to this characteristic bi-exponential relaxation decay in the FMs,  the relaxation process in AFMs is defined by only one, fast characteristic time, $\tau_{\rm{afm}}$. 
The relaxation time of the AFM order parameter $\tau_{\rm{afm}}$ is faster than $\tau_{1,\rm{fm}}$, for the same microscopic magnetic parameters. In particular, at low temperatures, the ratio between the relaxation times, $\tau_{1,\rm{fm}}/\tau_{\rm{afm}}$ is close to 5/3, like our predicted theory value for a sc lattice.
The absence of a second, slow relaxation process makes that the AFMs reach the final, equilibrium state much faster than in FMs, indeed the characteristic times,  
in FMs and AFMs, are related as $\tau_{2,\rm{fm}}\approx 12(1+4/z) \tau_{\rm{afm}}$, which ranges from 12 for $z\rightarrow{\infty}$ to 36 for $z=2$. 
As the final system temperature $T_{\rm{f}}$ approaches the critical temperature, $T_N$, the magnetization dynamics slows down both for AFMs and FMs. Figure \eqref{fig:relaxtimeEq}(b) shows the good agreement between our model (MFA-solid lines) and ASD (symbols) for both the AFMs and FMs.
The critical behaviour of the relaxation time can be also captured by a temperature scaling function, $\tau_{\|} \sim (1-T/T_{\rm{N}})^{-\nu}$ (dashed lines in Fig. \ref{fig:relaxtimeEq}(b)). By fitting our ASD simulation results, we find that for the FM system, $\nu_{\rm{fm}}=1.017(7)$, whereas for the AFM system,  $\nu_{\rm{afm}}= 0.64(2)$, in qualitative agreement with the prediction of our theory, the critical exponent in AFMs is smaller than in FMs.
We note that for the AFM fit, the critical exponent is obtained by taking only the data close to $T_{\rm{N}}$ into account, where the second term in Eq. \eqref{eq:effective-damping-AFM} dominates and therefore it coincides with our theoretical analysis. 
Another fundamental difference between FMs and AFMs is the dependence of the relaxation time on the number of neighbours $z$ to which each spin is coupled. Figure \ref{fig:relaxtimeEq}(c) shows the temperature dependence of the $\tau_{\rm{afm}}$ for three different lattice structures in 3D, sc ($z=6)$, bcc ($z=8$), and fcc ($z=12$), and in 2D, a square lattice ($z=4$). Lines in Fig. \ref{fig:relaxtimeEq}(c) correspond to the analytical estimation based in our model and symbols to the ASD simulations. We stress that the lattice structure dependence of $\tau_{\rm{afm}}$ only exists for AFMs. Relaxation time in FMs is independent of the lattice structure.

\begin{figure}[t]
\centering
\includegraphics[width=\columnwidth]{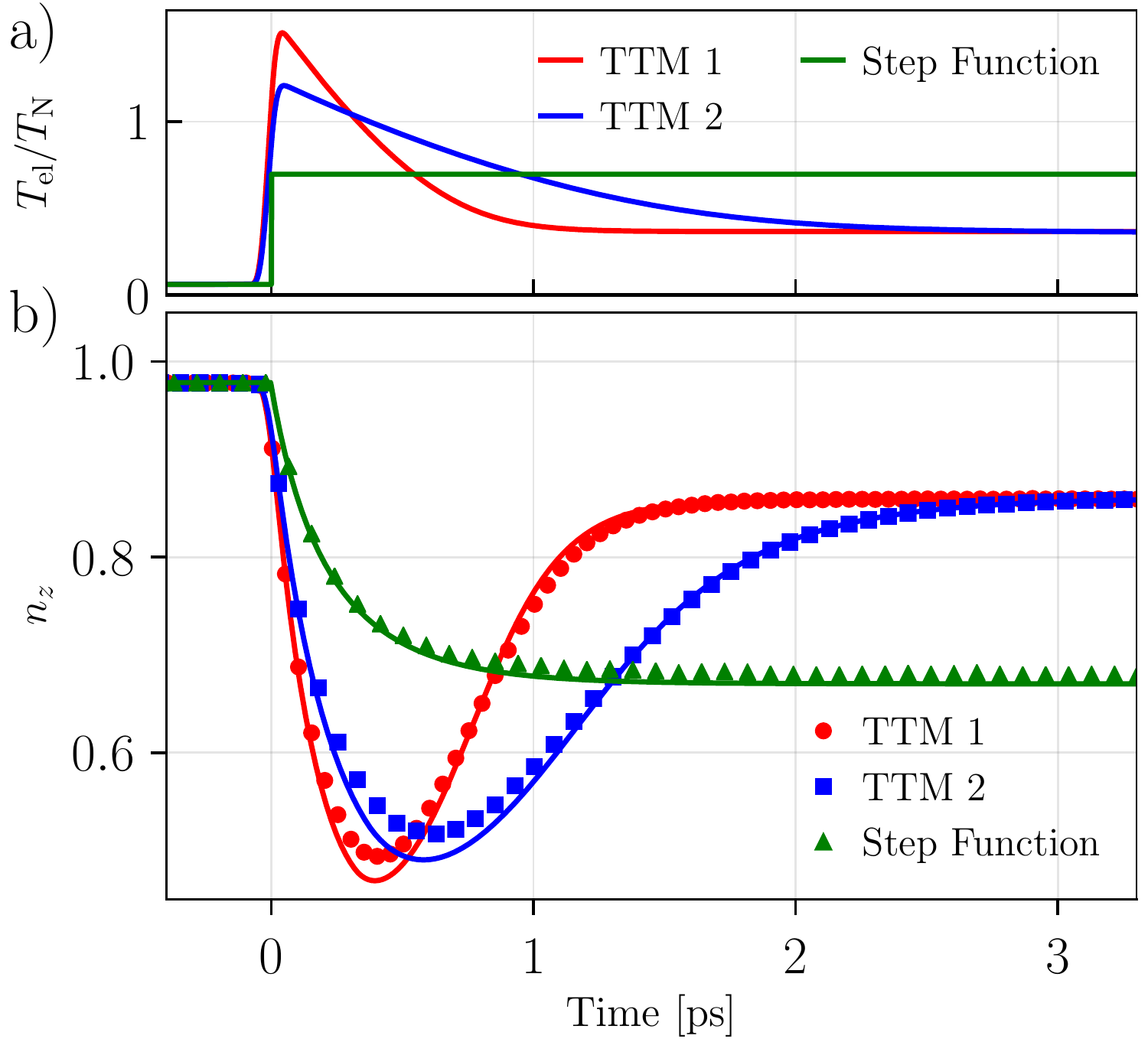}
\caption{(a) Temperature step function ($T_{\rm{i}}=0.06T_N$ and $T_{\rm{f}}=0.7T_N$) and two different $T_{\rm{el}}$ profiles for the same laser fluence and $g_{\rm{ep}}=6\times 10^{17}$J/sKm$^3$ for two  sets of parameters (TTM1: $\gamma=700$ J/K$^2$m$^3$, $c_\text{ph}=3 \times 10^6$ J/Km$^3$,  and TTM2: $\gamma=2000$ J/K$^2$m$^3$, $c_\text{ph}=5 \times 10^5$ J/Km$^3$). (b) The magnetic order dynamics as a response to the  temperature profiles in panel (a). The symbols correspond to ASD simulations and the lines to the numerical solution of Eq.~\eqref{eq:neel-dynamics}, ($\lambda=0.01$).}
\label{fig:AnalyticalvsSimulation}
\end{figure}

\begin{figure}[t]
\centering
\includegraphics[width=\columnwidth]{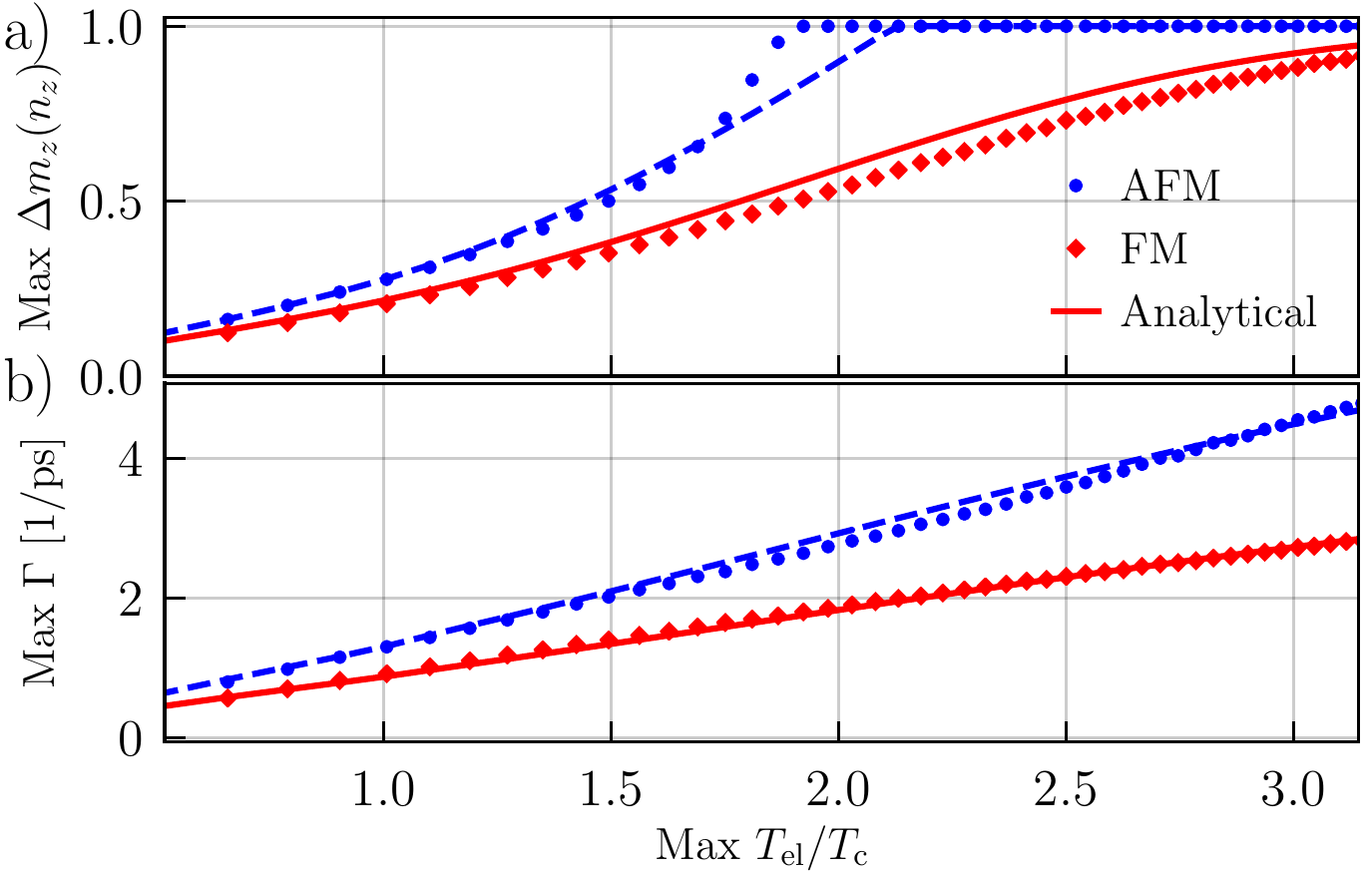}
\caption{(a) Maximum magnetic order quenching Max $\Delta m_z = \text{Max} |m_0 - m_z(t)|$ ($\Delta n_z = \text{Max} |n_0 - n_z(t)|$ for AFM) as function of the reduced peak electron temperature Max $T_{\rm{el}}/T_{\rm{c}}$. (b) Maximum demagnetization rate as function of the peak electron temperature.
Dots, AFM (blue) and FM (red), represent ASD simulations and dashed lines the numerical solution of Eq.~\eqref{eq:neel-dynamics}.
Results for a laser pulse of 50 fs duration and $\lambda=0.01$ in the ASD simulations.}

\label{fig:DemagvsLaser}
\end{figure}

One question remains, how could signatures of these exchange-enhanced dynamics in AFMs be found in experiments?
To address this problem, we first validate our model by comparing directly the dynamics of an AFM calculated via ASD  simulations and Eq. \eqref{eq:neel-dynamics} for three different temperature profiles (see Fig.~\ref{fig:AnalyticalvsSimulation}(a)). One temperature profile corresponds to a step function and the others to the TTM with two sets of parameters. The agreement between ASD simulations and our model is very good. We note that for quantitative comparison between ASD simulations and MFA models, one needs to slightly rescale the exchange parameter, $J$ (for more detail we refer to the supplemental material Sec. S2).
We find that Eq. \eqref{eq:neel-dynamics} describes ASD simulations as far as the microscopic spin configurations are homogeneous, as expected from the MFA grounds of our model (see details in supplemental materials, Sec. S3). 
By directly comparing the dynamics of FMs and AFMs under laser pulses, for instance see Figs. \ref{fig:Fig1} and \ref{fig:AnalyticalvsSimulation}, one can barely discern the effect of the exchange-enhancement in AFM dynamics. However, 
our model predicts striking differences between FMs and AFMs in the magnetic order dynamics in the limit of high-temperatures, $T\gg T_{\rm{c}}$, and small magnetic order parameter ($\xi=\beta J_0m \rightarrow{0}$). 
This scenario corresponds to experiments using powerful femtosecond laser pulses.
For FMs, Eq. \eqref{eq:long-LLB-1-ex} approximates to a linear equation: $(\mu_{\rm{at}}/\gamma)\dot{m} = 2 \lambda k_B T m$. Thus, the dynamics is described by an exponential decay, namely, it slows down as the magnetization $m$ reduces. In contrast, in the same limit, for AFMs, Eq. \eqref{eq:long-LLB-1-ex} approximates to $(\mu_{\rm{at}}/\gamma)\dot{n} \approx 4(4/z)  \lambda  k_B  T $, independent of $n$, which speeds up the AFM dynamics. This different dynamic directly emerges by increasing the laser fluence so that the electron temperature reaches very high temperatures and the magnetic reduces. In Fig. \ref{fig:DemagvsLaser}(a) one can observe the diverse behaviour of the maximum demagnetization ($\Delta_{\rm{max}} m(n)$) as a function of the reduced maximum electron temperature $T_{\rm{el}}/T_{\rm{c}}$ for both AFMs and FMs.
We note that since the results depend on the chosen TTM parameters, the results are drawn as a function of $T_{\rm{el}}/T_{\rm{c}}$ instead of laser intensity. Figure \ref{fig:DemagvsLaser}(a) shows that
for FMs the shape of $\Delta_{\rm{max}} m (T_{\rm{el}})$ is convex, while for the AFM, $\Delta_{\rm{max}} n (T_{\rm{el}})$ is concave.   
These findings align with an experimental work comparing the magnetic order dynamics of the AFM and FM phases in Dy~\cite{Thielemann-Kuhn2017}, where for comparable laser powers, the maximum demagnetization in AFMs was larger than in FMs~\cite{Thielemann-Kuhn2017}. 
It was also found that by increasing the laser intensity, the maximum demagnetization rate $\Gamma_{\rm{max}}$ increased in AFMs is much stronger than in FMs. 
Figure~\ref{fig:DemagvsLaser} (b) shows how the maximum demagnetization rate increases faster for AFMs than for FMs, in qualitative agreement with experiments conducted in Dy. The different slope of $\Gamma_{\rm{max}}$ is directly related to the exchange-enhancement of the effective AFM damping (Eq. \eqref{eq:effective-damping-AFM}), $1+4/z$, for sc used here, $\Gamma^{\rm{afm}}_{\rm{max}}=(5/3)\Gamma^{\rm{fm}}_{\rm{max}}$.

\textit{Summary.--} 
To summarize, we have shown that the ultrafast magnetic order dynamics in antiferromagnets is exchange-enhanced in comparison to ferromagnets with the same system parameters. The origin is the exchange-enhancement of the effective AFM damping. We have provided an equation of motion for the AFM magnetic order and predicted that AFMs have intrinsic faster dynamics and distinct critical dynamics than FMs. Notably, we have found that the exchange-enhancement strongly depends on the number of neighbours to which spins are exchange coupled, for instance in 2D magnets, the speed up of the dynamics is larger. In the very high temperature regime, we have predicted a transition from exponential to linear decay when the magnetic order reduces. We propose a method to discern this effect in experiments using powerful femtosecond laser pulses. We have demonstrate the validity of our model by direct comparison to atomistic spin dynamics simulations.

\textit{Acknowledgments.--}
We gratefully acknowledge support by the Deutsche Forschungsgemeinschaft through SFB/TRR 227  "Ultrafast Spin Dynamics", Project A08.

\bibliography{afm-lb}

\clearpage 
\widetext 

\setcounter{figure}{0}
\setcounter{page}{1}
\setcounter{equation}{0}
\renewcommand{\thepage}{S\arabic{page}} 
\renewcommand{\theequation}{S\arabic{equation}} 
\renewcommand{\thesection}{S\arabic{section}}  
\renewcommand{\thetable}{S\arabic{table}}  
\renewcommand{\thefigure}{S\arabic{figure}}

\textbf{Supplementary Material to "Exchange-enhancement of the ultrafast magnetic order dynamics in antiferromagnets"}

\section{S1. Two-temperature model} 

The dynamics of the electron temperature $T_{\rm{el}}$ and the phonon temperature $T_{\rm{ph}}$ can described via the two-temperature model (TTM) \cite{Kaganov1957,Chen2006}, 
\begin{align}
C_{\rm{el}} \frac{\partial T_{\rm{el}}}{\partial t} &= -g_{\rm{ep}}\left( T_{\rm{el}} - T_{\rm{ph}} \right) + P_{l}(t) \\
C_{\rm{ph}} \frac{\partial T_{\rm{ph}}}{\partial t} &= +g_{\rm{ep}}\left( T_{\rm{el}} - T_{\rm{ph}} \right).
\label{eq:2TM}
\end{align}
where $g_{\rm{ep}} = 6 \times 10^{17}$ J/sKm$^3$ is the electron-phonon coupling constant, $C_{\rm{ph}}= 3 \times 10^6$ J/Km$^3$ and $C_{\rm{el}} = \gamma_e T_{\rm{el}}$ ($\gamma_e = 700$ J/K$^2$m$^3$) represent the respective specific heats of the electron- and phonon system. Although we use standard values for metals, these values are material-dependent. $P_l(t)$ is Gaussian shaped and represents the absorbed energy of the electron system coming from the laser. 

\section{S2. Rescaling of the exchange constant for quantitative comparison between MFA and ASD simulations}

\begin{figure}[h]
\centering
\includegraphics[width=0.5\columnwidth]{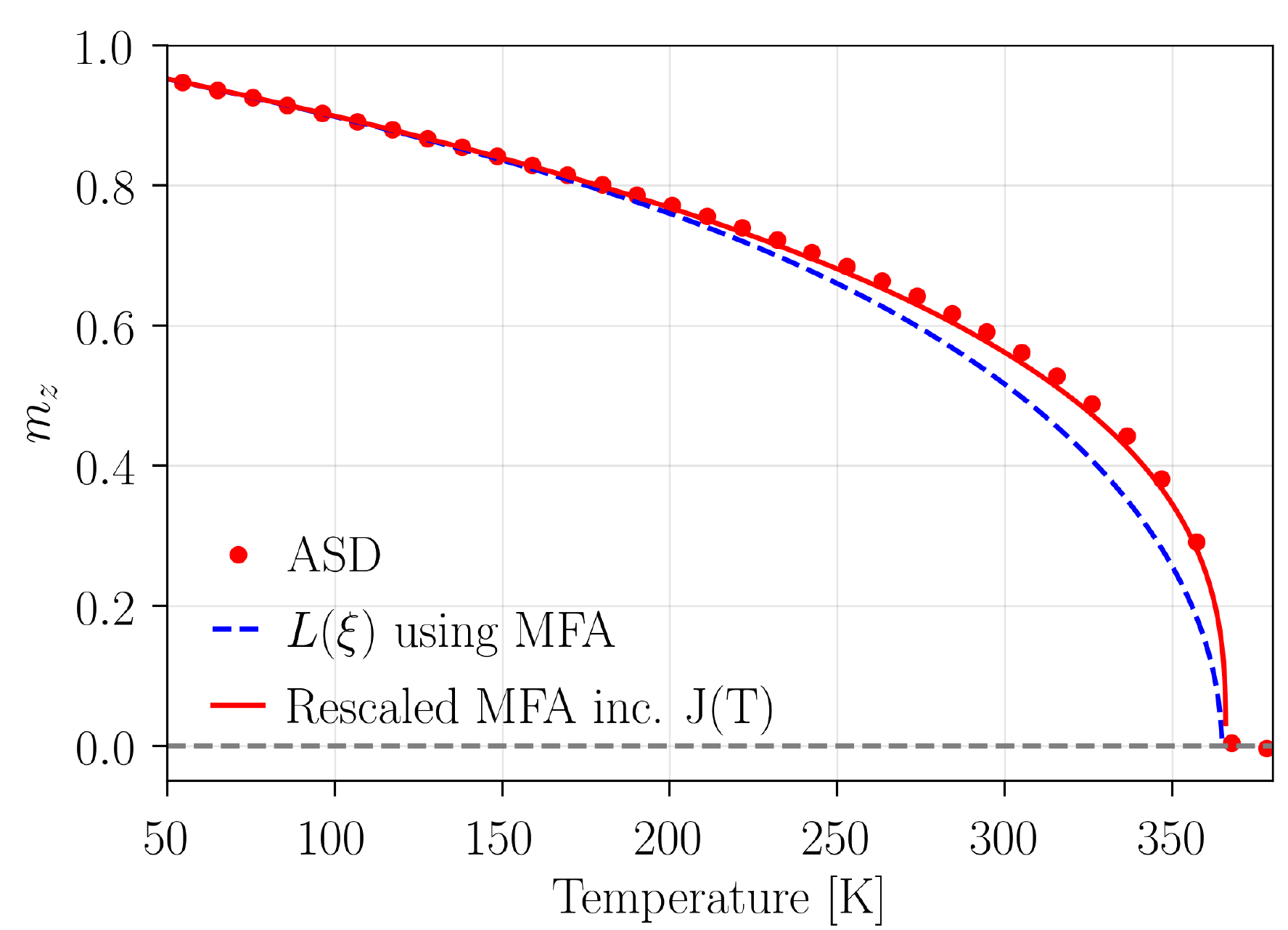}
\caption{Equilibrium magnetization of a sc-lattice as a function of temperature from ASD simulations (red dots), mean field approximation (blue dashed line) and from the MFA model including a temperature dependent rescaled Heisenberg exchange constant $J(T)$ (Eq. \eqref{eq:S2}), (red line).}
\label{fig:EquilibriumMag}
\end{figure}

In the main text, our analytical model for the magnetic order dynamics is based in the mean field approximation (MFA). The equilibrium magnetization as a function of temperature calculated using the MFA slightly differs from the ASD simulations.  Fig.~\ref{fig:EquilibriumMag} shows the MFA results as a blue dashed line and the ASD simulations as red points for a sc-lattice using $J= 3.450 \times 10^{-21}$ J.
For the MFA case, we have rescaled the exchange constant, $J_{\rm{mfa}}=0.73 J_{\rm{asd}}$, to obtain $T_N^{\rm{MFA}}=T_N^{\rm{ASD}}$. We have estimated the ASD critical temperature by calculating the temperature at which the magnetic specific heat diverge. 
The equilibrium magnetization as a function of temperature using the MFA start to deviate from ASD simulations in the intermediate-to-high temperature region, $T_N/2<T<T_N$. 
In order to quantitatively compare our model to ASD simulations we resolve this discrepancy by introducing a temperature dependent Heisenberg exchange modulation $J(T)=J_0 +J^\prime(T)$, where $J_0$ describes the original MFA Heisenberg exchange constant, $J_{\rm{mfa}}=0.73 J_{\rm{asd}}$, $J^\prime(T)>0$ is a temperature dependent modulation that needs to be determined. We determine it by forcing the equality the equilibrium magnetization calculated through ASD, $m_e=(1-T/T_N)^{1/3}$ ($1/3$ for a sc-lattice), and the MFA, $m_e= L (\beta J(T) m_e)$. Thus, the temperature dependent Heisenberg exchange $J(T)$ can be calculated from 
\begin{equation}
\left(1-T/T_{\rm{c}} \right)^{1/3}=L\left(\frac{(J_0 +J^\prime(T))m}{k_\text{B}T}\right)
\label{eq:S1}
\end{equation}
which can be solves as
\begin{equation}
J^\prime(T)= \frac{1}{\beta m} L^{-1}(\left(1-T/T_{\rm{c}} \right)^{1/3})-J_0.
\label{eq:S2}
\end{equation}

$L^{-1}$ describes the inverse Langevin function for which no analytical expression is known. However there have been numerous attempts at finding a simple and accurate approximation~\cite{Jedynak2015,Nguessong2014}. In this work we have used the equation proposed by Nguessong et al.~\cite{Nguessong2014} to approximate the inverse Langevin function numerically.\\
We note, that by using Eq.~\ref{eq:S2} $J(T)$ becomes independent of the numerical value of $J_0$ and is instead directly  calculated from the magnetization curve $m(T)$ via the inverse Langevin function. For a sc-lattice $m_e(T)=(1-T/T_c)^{1/3}$ agrees well with the atomistic results. However for other lattices (fcc, 2D or bcc), a different analytical expression for $m_e(T)$ is needed to describe $m_e(T)$.

\clearpage

\section{S3. Breakdown of the MFA model for high fluence laser excitation}
As discussed in the main text, our model is based in the MFA. This means that better agreement between ASD and MFA would be expected when the microscopic spin configurations remain close to the MFA assumptions, when each atomic spin sees the same interactions from the neighbouring ones. 
\begin{figure}[h]
\centering
\includegraphics[width=\textwidth]{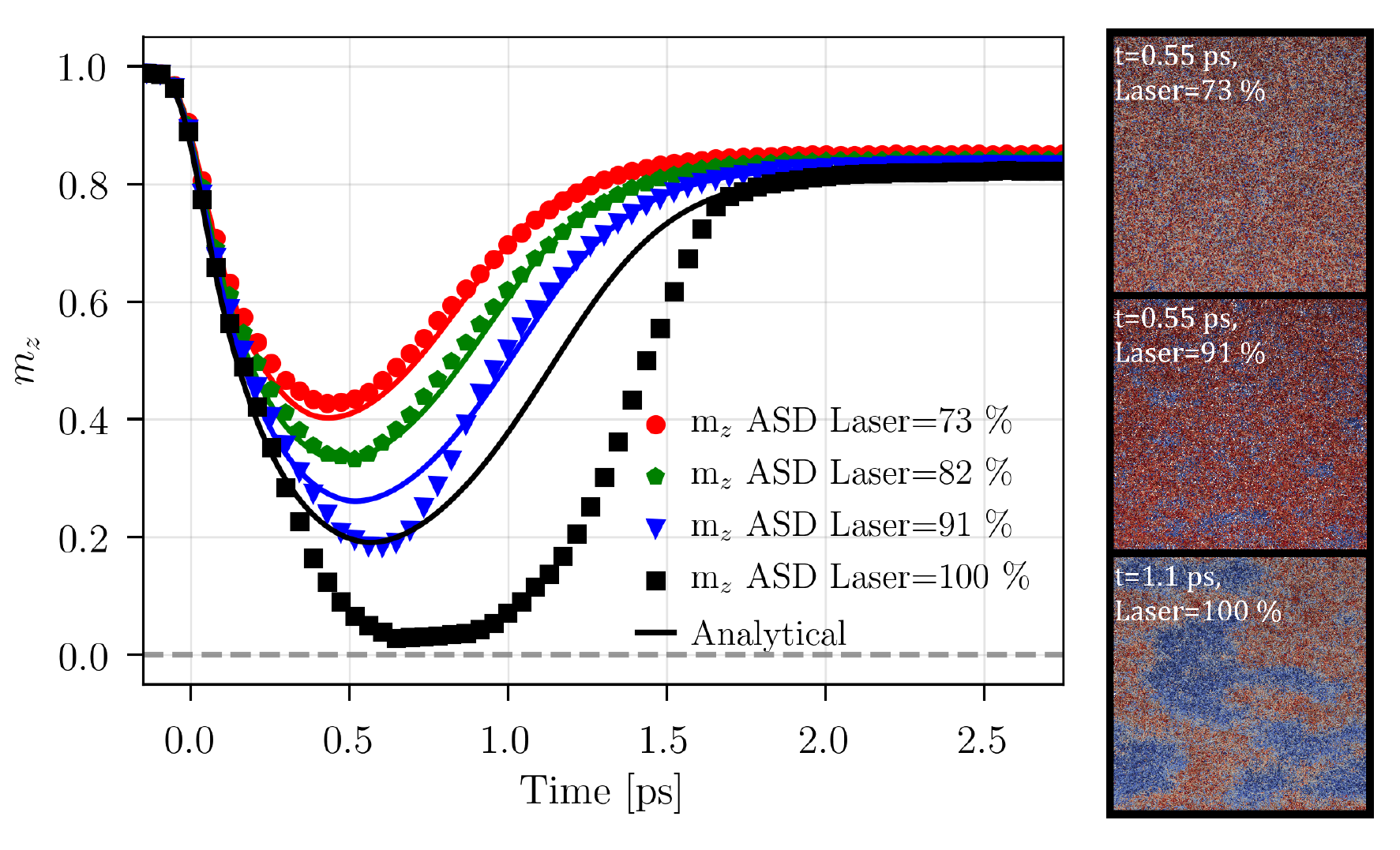}
\caption{
ASD simulations of AFM magnetic order dynamics for different laser powers ($\lambda=0.01$) (dots) in comparison to our analytical model (Eq.~\eqref{eq:neel-dynamics}) (lines). Higher laser powers yields larger demagnetization and the underlying MFA assumptions of Eq.~\eqref{eq:neel-dynamics} stops being a valid approximation. On the right different states of the spin system are shown, shortly after the excitation with a laser pulse.}
\label{fig:Domain}
\end{figure}

When magnetic domains are be nucleated, our MFA macroscpin model no longer describes the spin state correctly. Figure~\ref{fig:Domain}. shows the magnetic order dynamics for a range of laser fluences, where symbols correspond to ASD simulations and lines to the macrospin model. For higher fluences the agreement between the two models diminishes.
The right side shows snapshots of the microscopic spin configuration at different time delays corresponding to the a time range where maximum demagnetization is achieved. When the laser fluence is only the 73 $\%$ of the maximum fluence simulated, the microscopic spin configuration is homogeneous. In that case, the agreement between theory and simulations is very good. As the laser fluence increases, magnetic domains start to nucleate and the theory and simulations to deviate. For the maximum laser fluence that we simulate 100 $\%$, large magnetic domains are nucleated and the MFA breaks down. The theory is not able to describe this situation. For those cases, a micromagnetic model should be developed.

\end{document}